# NON-DESTRUCTIVE MECHANICAL CHARACTERIZATION OF SiC FIBERS BY RAMAN SPECTROSCOPY.

Gwénaël Gouadec[a,b] and Philippe Colomban[a,∂]

**(a)** Laboratoire Dynamique-Interactions-Réactivité (UMR7075-CNRS & Université Pierre et Marie Curie), 2 rue Henri Dunant, 94320 Thiais, France.
**(b)** Office National d'Etudes et de Recherches Aérospatiales (ONERA), Département Matériaux & Systèmes Composites (DMSC), BP 72, 92322 Chatillon, France.

*ABSTRACT*

The paper provides a comprehensive study on Raman spectroscopy versatility as a fast and non destructive tool for the prediction of the mechanical properties of SiC fibers derived from a polymeric precursor (NLM™, Hi™, Hi-S™, SA™ and Sylramic™ grades) or produced by CVD (SCS-6™ fiber), including in situ analysis in CMCs or MMCs. We show how the results of very simple spectra fitting are correlated with Young's modulus, tensile strength and microhardness. The reason why such a correlation exists, the common dependency of Raman signal and mechanical behavior to the micro/nanostructure of ceramics, is discussed.

---

[∂] Author to whom correspondence should be addressed
Fax : 33 (0) 1 49 78 13 18
e-mail : colomban@glvt-cnrs.fr

# 1 Introduction

The reinforcement of ceramic materials by long ceramic fibers leads to low density and refractory materials, of high damage tolerance, that should be appropriate for metal alloys substitution in advanced engines (turbines) and waste treatment energy plants[1]. Ceramic fibers can also be incorporated directly in metal matrices to increase their high temperature mechanical properties[1].

SiC fibers, which are among the most stable fibers, have always been produced by the 3D reticulation of a polymeric precursor. This synthesis route leads to a high matter homogeneity and very smooth surfaces, the lack of defects explaining tensile strengths **$\sigma_r$** as high as 3 GPa. SiC fibers have been widely studied, especially their aging[2-7] which evidenced that densification is linked to compositional changes (for instance the loss of some residual hydrogen[8]). A peculiarity of ceramics issued from polymeric precursors is no impurities are concentrated at the grains boundaries. Abnormal grain growth is very common in the lack of such usual diffusion regulators[9] and the know-how of SiC fibers manufacturers consists in postponing the onset of SiC crystallization and grain growth, since a high correlation is anticipated with mechanical degradation. A close relationship has been evidenced, for instance, between the micro-hardness and short-range ordering in sol-gel prepared nanocrystalline oxides[10] and some authors have already pointed out fibers overall mechanical ability is governed by their microstructure[5,7,11].

The purpose of this paper is to try correlating, to the best extent possible, the tensile strength ($\sigma_r$), the Young's modulus (**E**) and the micro-hardness ($\mu H$) of different SiC fibers, either to Raman spectra or to "Raman Extensometry" S coefficient (measuring the strain-induced shifts of Raman bands). Unlike most experimental methods intended to measure directly $\sigma_r$ and **E**, Raman analysis does not require extracting fibers from the matrix, either by mechanical grinding/crushing[12] or by chemical attack[13]. Only the most resistant fibers are extracted in the former case while fibers surface might be altered in the latter. Besides, it will always be difficult to extract long enough fragments to carry tensile strengths measurements and getting representative results will not be trivial. Thus, provided the matrix is sufficiently transparent, non destructive *in situ* Raman analysis is possible through matrix thickness' of 5-30 µm. It is a tool for composition and structural investigation, with a high sensitivity to the chemical bond nature and the short range arrangement in the first (0.1-0.5 nm) and second ($\leq$0.5-5 nm) atomic shells, whatever the crystallinity level (including amorphous nanophases[2,14-17]). Besides, Raman spectroscopy is not demanding in terms of sample preparation and relatively rapid to process. It has already been used to follow the chemical alteration of SiC fibers under thermal treatments[18].

# 2 Experimental

## 2.1 Raman equipment

All Raman spectra except those of fig. 6(a) and carbon spectra of fig. 6(b) were recorded with a "XY" spectrograph (Dilor, France) equipped with a double monochromator as a filter and a back-illuminated liquid nitrogen-cooled 2000 x 800 pixels CCD detector (Spex, a division of the Jobin-Yvon Company, France). The spectra were recorded in "microconfiguration" (laser beam focused through a microscope to a 1µm diameter impact on the sample) and "backscattering setting" (backwards collection through the microscope). We used the 514.5nm "green" line of a "Innova 70" Argon-Krypton laser source (Coherent, USA) and the power was measured under the microscope objective thanks to a "PD200" photodiode detector (Ophir, U.S.A.). This power was kept in the range 2-4 mW since too large a difference can induce thermal disturbances to the signal[19,20].

The spectra of fig. 6(a) as well as carbon spectra in fig. 6(b) were recorded with the 632.8nm line of an He/Ne laser on a "System 1000" spectrometer cooled down to 200K by Peltier effect (Renishaw, U.K.).

## 2.2 Tested fibers

Room temperature Raman spectra were recorded on cross sections of **i)** NLM™, Hi™ and Hi-S™ grades of Nippon Carbon (Japan) Nicalon fibers ; **ii)** the SA™ fiber of UBE Industry (Japan) ; **iii)** the Sylramic™ fiber from Dow Corning (U.S.A.) and **iv)** the SCS-6™ fiber from Textron (U.S.A.). The FT700™ carbon fiber from Tonen (Japan) will also be considered, for discussion purposes. The properties of the fibers are given in table 1. The reason why core examination was preferred to a surface analysis is a lower structural variability.

## 2.3 Samples preparation

Some NLM and Hi fibers were annealed in oxidizing and reducing atmospheres before Raman analysis[2]. The thermal treatments were assumed to be equivalent to those performed by Berger et al[21] and Kumagawa et al[22] on, respectively, Hi and NLM fibers the strength of which they measured.

A fiber polishing was required before mapping the cross-section of a raw SCS-6 fiber. On account of its large diameter and its composite structure, with strong differences in hardness between the carbon core and the different SiC layers, a nickel coating had been deposited by an electrochemical method before resin embedding and polishing. We checked this nickel deposit did not provoke any chemical and/or mechanical alteration[23].

## 2.4 Spectra treatment

The fitting of experimental carbon[19] and SiC[24] spectra has been described elsewhere. Briefly, the first step includes the subtraction of a linear baseline attached to the spectra limits. Then, bands are then preset on the basis of reported spectra contributions and are attributed either lorentzian or gaussian forms. The fitting software eventually produces the wavenumber ($\bar{\nu}$), the width (FWHH, which stands for full width at half height) and the intensity of each band. Two examples of fits are given in fig. 1.

## 3 Results

### 3.1 Correlation between Raman spectra and microstructure

For Raman spectra to be related to the mechanical properties, one must identify the fingerprints of all phases and assign their components to given bonds. Fig. 1 presents the spectra recorded on fresh fiber sections. The Raman scattering efficiencies of graphite and SiC are in a ratio of about 20 which means a direct intensity comparison does not give the respective proportions[2].

In low crystallinity solids like "SiC" fibers, carbon spectrum mainly consists of a strong "$sp^3$-like" resonant contribution and a "$sp^2$-like" doublet[2], as observed in carbon containing films[25-30]. The former is fitted by a lorentzian band peaking between 1325cm$^{-1}$ (under red excitation) and 1365cm$^{-1}$ (blue excitation). As for $sp^2$ components, they are best fitted by two gaussians set at 1590 and 1625cm$^{-1}$ (whatever the excitation ; note the second gaussian is absent in the NLM and Hi fibers, only a shoulder in others and will therefore not be in the scope of this study). The above-mentioned modes are stretching modes and assignments are fully discussed elsewhere[31]. Another band around 1530 cm$^{-1}$ must be taken into account in the NLM and Hi grades and was attributed to C-C bonds in contact with heteroatoms (e.g.

oxygen). It is admitted that a carbon observed with the 514.5nm line is graphitic from the moment when intensities $I_{sp^3}$ and $I_{sp^2}$ become equivalent[2]. Otherwise, carbon is diamond-like rather than graphitic, the tetrahedral coordination offering a better structural compatibility with the $sp^3$ hybridized atoms of silicon carbide network.

The SiC spectrum is not detectable in the NLM and Hi grades on account of their carbon excess (atomic C/Si ratio is respectively 1.3 and 1.4 ; table 1). Only in the very last generation of SiC fibers can SiC spectrum be recorded (Hi-S, SA and Sylramic fibers all have a C/Si ratio below 1.1) and it mainly consists of peaks around 796 and 972 cm$^{-1}$. These peaks correspond respectively to the transversal optic (TO$_1$ and TO$_2$ ; degenerate) and the longitudinal optic (LO) modes of β-SiC (cubic structure, 3C-SiC in Ramsdell notation), which was identified by x-rays diffraction as the SiC phase present in Nicalon (NLM, Hi & Hi-S)[32], SA[22] and Sylramic[33] fibers. Yet, satellite Raman lines like the prominent contribution around 765 cm$^{-1}$, are detected. They result from SiC polytypism, when Brillouin Zone is folded because of a regular faulting in β-SiC stacking of carbon and silicon layers[34]. A broad background assigned to a so-called "Density of State" contribution superimposes between 700 and 1000 cm$^{-1}$. Such phenomenon arises from "random" faulting in the stacking sequence[2].

Fig. 2(a) compares the effect of an oxidizing thermal annealing on Raman FWHH of "sp$^3$-carbon" mode for the three grades of Nicalon fibers. Fig. 2(b) shows some of the corresponding core spectra for the NLM fiber.

### 3.2 Comparison of Raman spectra with fiber strengths

Fig. 3(a) is a double scale plot of Hi-Nicalon strength and Raman spectra dependency to annealing, from room temperature up to about 1500°C. The Raman parameter is the FWHH of "sp$^3$-carbon" band and was obtained on spectra recorded at room-temperature on the surface of annealed fibers. The strengths are those measured during thermal treatments by Berger et al[21]. In fig. 3(b), each point corresponds to a given annealing temperature and a Raman parameter, either the wavenumber or the FWHH of "sp$^3$-carbon" band, is given as a function of the high-temperature strength. The strength values for the annealing temperatures of the Raman investigated fibers were extrapolated from a sigmoïdal fit of Berger et al data[21] (Hi-Nicalon) and a exponential fit of Kumagawa et al data[22] (NLM-Nicalon).

### 3.3 Young's modulus and "Raman microextensometry"

*In-situ* measurement of composites strain by "Raman Extensometry" uses calibrations of the wavenumber shift ($\Delta \bar{v}$) induced on a fiber stretching mode by a tensile strain ($\Delta \varepsilon$) :

$$\Delta \bar{v} = S \times \Delta \varepsilon \qquad (1)$$

S is usually negative and a compression (respectively a tension) increases (decreases) the wavenumber. The method is already used on model Carbon/Epoxy composites[35-40] and applicability to ceramic- and metal-matrix-composites has been demonstrated recently[41-47] despite the high consolidation temperatures which can induce significant structural evolution (by fiber/matrix interdiffusions) and complicate the study.

Even though bonds anharmonicity is usually considered as a simple perturbation to the harmonic case, it has a very significant influence on the physical behavior of materials. As a matter of fact, only anharmonicity can explain thermal expansion or the finite value of thermal conductivity[48] and the same goes for the Raman shifts of equation (**1**). The vibration frequency of a "bond-simulating" harmonic spring is indeed proportional to $\sqrt{k}$, where **k** is the so-called "harmonic force constant" and does not depend on the bond length (**l$_b$**). In first approximation, the simplest expression modeling a "realistic" bond potential **V**(l$_b$) is the following (l$_0$ is l$_b$ equilibrium value) :

$$V(l_b) = \frac{k}{2}(l_b - l_0)^2 + k'(l_b - l_0)^3 \quad (2)$$

where k' accounts for bonds anharmonicity. If we consider, by analogy with the harmonic case, that vibrational wavenumbers $\bar{v}$ are proportional to the square root of $\frac{\partial^2 V}{\partial l^2}$, then :

$$\bar{v}(l_b) \propto \sqrt{k \times \left(1 + \frac{6k'}{k} \times (l_b - l_0)\right)} \quad (3)$$

$\bar{v}(l_b)$ has the form $\sqrt{k} \times \sqrt{(1+x)}$. k' is expected to be much smaller than k so, as long as $l_b$ remains close to $l_0$ ("small strains approximation"), a Taylor expansion is justified for $(1+x)^{1/2}$ and :

$$\bar{v}(l_b) \propto \sqrt{k} + \frac{3k'}{\sqrt{k}} \times (l_b - l_0) \quad (4)$$

$$\bar{v}(l_{b_2}) - \bar{v}(l_{b_1}) \cong \frac{3k'}{\sqrt{k}} \times (l_{b_2} - l_{b_1}) \quad (5)$$

By comparison with equation (**1**), one eventually finds that S is proportional to $k' \times k^{-1/2}$. Besides, according to the "bond compression model", Young's modulus **E** in isotropic structures must be linked by proportionality to the force constants of all bonds present in the material[49] :

$$E = \sum_{bonds} \frac{l_b^2 k_b}{9} \quad (6)$$

If so, S should eventually be proportional to $k'E^{-\frac{1}{2}}$. Fig. 4 groups Raman bandwidths ("sp$^3$" mode) for several fiber grades, plotted as a function of Young's Modulus measured at room temperature. Fig. 5 shows S absolute values collected on the literature work (from our group[50] and others[11,36,39,40,51-64]), as a function of $E^{-1/2}$.

### 3.4 Raman spectra and local microhardness (µH)
The SCS-6 fiber is a 140µm diameter fiber used for metal-matrix reinforcement[1]. It has a high heterogeneity along the radius, due to a preparation by CVD of SiC (and carbon) on a carbon core[65]. The resulting variations in "Berkovich's µH" (three-sided diamond tip) and Young's modulus could be measured by Mann and coworkers[66]. Fig. 6 was drawn to find out whether such variations (fig. 6(c)) could be related to Raman spectra. In fig 6(a), fitted band intensities have been summed for the three carbon peaks, on the one hand, and SiC optical modes (TO & LO) on the other hand. As for fig. 6(b), it represents the FWHH of one SiC and one carbon peak. Note SiC points correspond to three separate scans and evidence a good reproducibility of the measurements.

### 4 Discussion

### 4.1 Raman spectroscopy and SiC fibers micro/nano-structure
Covalent materials can be described on the basis of a "molecular approach" (as opposed to a "crystal symmetry" approach) and each stretching mode is then specific to one given chemical bond. The bond wavenumber will be a function of its stiffness, of its length and of the mass of its atoms while the bandwidth will reflect the short range disorder (static or dynamic, topologic or electric). Both the wavenumber and the bandwidth should therefore correlate with mechanical properties, as discussed in next paragraphs.

Raman bands being a projection on the energy axis of the configurations distribution, crystallization of amorphous phases and subsequent grain-growth drastically reduce the bandwidths. Concerning carbon Raman spectra, resonance allows for $sp^3$ hybridization specific probing[25,27,67,68]. Comparison of $sp^3$ and $sp^2$ hybridization in fibers from different grades becomes possible. Structural changes during thermal treatments[2] or matrix-embedding[2,45] can be followed and the spectra of fig. 1 & fig. 2 are obviously correlated with the microstructural data of table 1. First, the presence of the 1530 cm$^{-1}$ band is obviously an indication of the oxygen content. Indeed, this band has a strong intensity in the NLM fiber (12 wt% O), which reduces in the Hi-Nicalon (0.5wt%) and vanishes in the three nearly stoichiometric SiC fibers. The oxygen is known to govern creep resistance and room temperature Young's modulus (oxygen is incorporated into a "soft" oxycarbide forming a continuous network in the fiber structure). NLM Raman spectrum has a strong shoulder on the low frequency-side of the "$sp^3$-carbon" band, which is best seen under red excitation[69]. It must be related to the hydrogen that was shown to surround carbon precipitates[70] (C-C vibrations disturbed by C-H bonding). The band narrowing observed in the Hi-Nicalon with respect to NLM spectra is consistent with a greater size of carbon precipitates and an increasing organization. We do not have carbon precipitates size in the nearly stoichiometric fibers but it becomes possible to resolve the graphitic doublet. Carbon was shown to lie on SiC grains in the NLM[6] and Hi[5] grades. In the SA fiber, carbon is found at grain boundaries triple points[21], which implies it has strongly evolved and must be well organized, in rather big entities. Fig. 1 shows SA & Sylramic fibers have similar SiC spectra. The Hi-S one differ by a stronger polytype contribution, which suggests a lower short range order, in agreement with the 10nm size of SiC microcrystals ($\geq$ 100nm in the SA and Sylramic grades). In addition, the fitting of Hi-S silicon carbide spectrum shows $TO_1$ and $TO_2$ degeneracy is lifted. This would not occur for pure 3C-SiC and LO mode would peak around 972 cm$^{-1}$ instead of 966cm$^{-1}$. Sasaki et al concluded in favor of a 2H (hexagonal) polytype contribution in heat-treated NLM fibers[18] whereas the four peaks labeled for the as-received Hi-S fiber in fig. 1 rather correspond to 6H (hexagonal) or 21R (rhomboedric) polytypes[71].

The NLM, Hi and Hi-S grades of Nicalon fibers withstand annealing up to 1200-1300°C[18], 1500°C[72] and $\geq$1600°C[32], respectively. The low stability of the NLM fiber is attributed to the internal oxidation of its "free" carbon by the oxygen of the oxycarbide phase[18,73]. As for Hi and Hi-S fibers degradation, it results from a SiC oxidation by "atmospheric oxygen", which becomes fast as soon as the passive silica scale that forms around 900°C[74,75] is degraded[22,73,75,76]. It is obvious however looking at fig. 2(a) that the NLM and Hi fibers spectra start evolving from 1100 and 1250°C (Hi-S spectrum is stable up to 1400°C but we have no data above). Thus, the slope changes in fig. 2(a) detect the beginning of the fiber degradation mechanisms before they actually affect the properties. The fibers retain good properties until all carbon has disappeared (carbon acts as a grain growth inhibitor for SiC[18] and restricts diffusion[74]) and Raman spectroscopy can be used to find fibers stability limits. As for fig. 2(b) spectra, they show the bonds reorganization when the low density and amorphous "skeleton" of the NLM fiber, issued from a fired polymer (polycarbosilane), is transformed into a fully dense crystalline ceramic. The progressive shift of "$sp^3$-carbon" band from ~1340cm$^{-1}$ (for 1000°C heated fibers) to ~1355 cm$^{-1}$ (when temperature has reached 1500°C) shows the "aromatization" of carbon nanoprecipitates : a conversion of simple C-C bonds into C$\stackrel{...}{=}$C aromatic bonds. This is correlated with the increase of the direct current electronic conductivity[45]. The simultaneous band narrowing arises from crystallization onset and the internal reaction between the oxycarbide and free carbon second phases. Scanning electronic microscopy evidenced SiC crystals also develop on the fiber surface (unpublished results). When the carbon content, hence light absorption, decreases sufficiently, SiC Raman fingerprint can be observed.

Once "aromatization" is achieved, above 1500°C, Van der Waals interactions between the graphitic planes that have formed in carbon nanoprecipitates become more and more intense. Ordering takes place in the stacking direction with a small downshift of the 1590cm$^{-1}$ band[2]. The intensity increases simultaneously and becomes higher than that of the "sp$^3$" mode.

### 4.2 Strength prediction from Raman wavenumber and bandwidth

Fig. 3(a) evidences a perfect correlation exists between the Raman spectra of SiC fibers and their strength. This is also true for another mechanical property, "Vickers's µH", as was reported by Gogotsi et al for semiconductors, quartz and carbon[15,17] and by Amer et al for "diamond-like carbon"[16].

Strength can be described as a summation of chemical bonds responses to local stresses and Raman bands are, precisely, characteristic of chemical bonds. The vanishing of fig. 3(a) curves starts at very close temperatures, which confirms the overall macroscopic behavior of the fibers is governed by their microstructure : the above-mentioned silica layer no longer protects the fiber beyond 1200°C. Besides, the linear relationship in fig. 3(b) between Raman features (wavenumber for the NLM grade, bandwidth for the Hi one) and strength suggests microscopic and macroscopic responses obey the same phenomenon over the whole tested temperature range. A strength reduction resulting from grain boundary creeping effects is therefore ruled out since they would not contribute to Raman spectra.

On the basis of fig. 3(b), the typical strength sensitivity of carbon "sp$^3$" mode would be 15 cm$^{-1}$/GPa for the wavenumber and 10 cm$^{-1}$/GPa for the bandwidth. Such linear trend was encountered for "sp$^3$" peak area in the above-mentioned study on diamond films hardness[16]. Two atmospheres were tested in the case of the NLM grade and the mechanical failure postponing in the case of the reducing atmosphere does not seem to modify the so called "micro-macro" mechanical correlation.

### 4.3. Raman spectra relationship with Young's modulus and the micro-hardness

Of course, the proportion and the distribution of carbon and SiC phases in the different fibers determines separate "families". If we consider carbon in the fibers used for fig. 4, it changes from a dissolution in a "soft" oxycarbide (NLM) to turbostratic nanoprecipitates (Hi), which organization increases (SA, Sylramic, Hi-S) until carbon becomes the only phase present (almost fully crystalline and parallel planes of the FT700 fiber). The FWHH gap between the NLM fiber and the other grades results from its higher oxygen content. On the other hand, FT700 high modulus comes from the alignment of graphitic planes along the fiber axis. Note FT700 has almost the same FWHH as the Hi-S fiber, in spite of the long-range alignment of its graphitic planes. However, the difference does not concern the first shell environments (the short-range order, under 5nm, is similar in the graphitic domains of both fibers) and has therefore no effect on Raman FWHH.

Contrary to fig. 4, fig. 5 allows for a global classification of all carbon-containing fibers. The scales chosen for the representation lead to a linear positioning, in agreement with the discussion of equations (1) to (5). Yet, deviations to the general trend cannot be fortuitous :

**i**)- The reasoning that drove to a $E^{-1/2}$ abscissa scale supposed anharmonic effects had the same importance whatever the sample. k' is likely less sensitive than k to the system but it might change however. This is probably the case in SiC fibers (Tyranno and NLM grades) where carbon is only a second phase. **E** must be significantly affected by Si-C bonds and there is no reason why the $\frac{k}{k'}$ ratio of such bonds should be the same as that of C-C bonds.

Further work on S coefficients based on SiC spectra as a strain probe would be very interesting to compare with fig. 5 results.

**ii**)- Equation (5) best describes the "small strains domain" whereas S coefficients are measured till fiber failure. In fact, some authors even replaced the linear dependency of equation (1) (hence that of equation (5)) by a $\{a\Delta\varepsilon + b\Delta\varepsilon^2\}$ polynomial law, so as to take into account a reported bond hardening in compression (atoms repel each other) and the reciprocal softening in tension (bonds tend to dissociate). Tarantili et al found a/b typically equal to 25 for the "symmetric" C-C stretching mode of polyethylene fibers[77] but Melanitis et al found a/b~5 (only !) in "PAN-based" carbon fibers[78]. However, SiC fibers being much more isotropic than carbon fibers (strong covalent bonds in all directions), a greater reversibility is expected when passing from tensile to compressive stress. Hence, lower deviations to linearity and b should be negligible.

**iii**)- Carbon is a resonant species, which means its spectrum depends on the wavelength. This has been known for a long time from a "spectroscopic" point of view but papers on Raman extensometry of carbon have always neglected this aspect of the problem. We did find preliminary results suggesting S coefficients are a function of the laser wavelength and all S values in fig. 5 were not obtained with the same line.

In conclusion, fig. 6 illustrates in a convincing way what we had in mind writing this article. At first sight, every "mechanical change" visible in fig. 6(c) is related to a change either on carbon spectrum or on SiC spectrum (fig. 6(a) & fig. 6(b)). The correspondence has been highlighted by vertical dots delimiting zones that we numbered in Roman font. From core to periphery, they correspond to :

**I**)- the 32µm diameter carbon core consisting of graphitic units 1 to 5 nm in size[23].

**II**)- a 1.5 µm thick layer of pyrolytic carbon (grains are 25-50 nm in size). It has been identified as the weakest part of the fiber because of graphitic planes being parallel to the fiber axis[66].

**III**)- a zone in which carbon coexists with SiC and evolves over 30 µm towards a highly disordered carbon ("$sp^2$" and "$sp^3$" bands of carbon Raman spectrum broaden, while the "1530 $cm^{-1}$" one gets stronger[23]). SiC contribution is made abnormally weak by carbon efficiency an order of magnitude greater but the way the three series of fig. 6(b) get closer and closer illustrates SiC better organization and the hardening (β-SiC grains grow from 12 nm to 100nm[66]).

**IV**)- a zone where only SiC is detected (C-C bonds abundance $\leq 0.1\%$). TO and LO peaks widening and the increase of the background between them reveals stacking faults are all the more frequent as thickness increases[23]. In spite of this, mechanical properties are constant, which suggests they are governed by free carbon in insertion (while Mann et al attributed the high variations in zone II to SiC structural evolution[66]).

**V**)- the interfacial carbon deposit (thickness 3µm), with a huge fluorescence contribution[23]. Due to Raman extremely high sensitivity to carbon, we could detect some free carbon in the very last microns of SiC outer sheet. We guess they must explain the slight edge degradation observed in fig. 5(c).

One point to note is that the five regions do not correspond at all to the regions that can be identified "optically". Note also that although **E** determination is hard to perform at high temperatures, it would probably be a better parameter to follow for macro-mechanics characterization than the strength (used in fig. 1 & fig. 2), which is highly dependent on the tested sample.

## SUMMARY - PERSPECTIVES

Raman spectroscopy being a vibrational analysis, it had to depend on micromechanics. This technique actually allows for studying annealing impact on the mechanical degradation of SiC fibers and helps understanding the mechanical properties (E, $\sigma_r$ and µH) of different fiber grades, on the basis of chemical considerations. Special emphasis was put on the relationship between Young's modulus on the one hand, in other words the macroscopic response to applied stresses, and the so called S coefficient of "Raman extensometry", which depends on bonds vibrations and is therefore a micromechanical parameter.

Resonance makes laser penetration in carbonaceous materials dependant on the wavelength, since absorption is not constant. This makes us expect the possibility of measuring stresses as a function of the distance to the surface. Yet, resonance still has to be assessed since Galiotis & Batchelder (cited in Amer et al[16]) gave an estimate 60 µm laser penetration in carbonaceous materials when comparison of Raman spectra in SiC fibers and amorphous carbon-rich SiC films suggests penetration could be lower than 100nm using blue-UV exciting lines[20]. Further work will require the precise assignment of carbon bands as well as S measurements for SiC phase. Besides, measurements of S coefficients under lasers polarized parallel or perpendicular to the fiber axis could indicate grains preferential orientations, if the results were different.

**CAPTIONS (Table 1. and figures)**

**Table 1.** Properties of commercially available SiC fibers.

**Fig. 1.** Carbon (left) and SiC (right) spectra recorded on the successive generations of polymer-derived silicon carbide fibers ($\lambda$=514.5nm ; P=2mW ; t=90s, unless specified).
One fitting is given in example in each case.

**Fig. 2. (a)** Full Width at Half Height (FWHH) of "$sp^3$" carbon mode as a function of thermal treatments performed in air for SiC fibers synthesized by oxygen (NLM) or electronic (Hi, Hi-S) curing of a polymeric precursor (polycarbosilane) ; **(b)** Changes of NLM core spectra on annealing in a reducing atmosphere.
Raman spectra in **(a)** and **(b)** were recorded with the 514.5nm line.

**Fig. 3. (a)** Comparison for different temperatures of Hi-Nicalon fibers tensile strength[21] (black circles) with the Full Width at Half Height (FWHH) of their "$sp^3$-carbon" stretching mode on surface recorded Raman spectra (white circles) ; **(b)** wavenumber and FWHH of "$sp^3$-carbon" mode plotted as a function of tensile strengths found in Berger et al[21] or Kumagawa et al[22] for heat treated fibers.
Raman spectra in **(a)** and **(b)** were recorded with the 514.5nm line.

**Fig. 4.** Raman bandwidth ("$sp^3$" C-C mode) and Young's Modulus of as-received SiC fibers. FT700 carbon fiber (Textron, U.S.A.) is presented for comparison.

**Fig. 5.** Absolute value of the stress sensitivity coefficient (S) of carbon "$sp^2$" band plotted versus the inverse of Young's Modulus square root.
**References :** Kevlar™(poly(p-phenylene terephtalamide))[11,52-54] ; aramid fibers other than Kevlar™[11,40,51] ; PAN-based carbon fibers (IM43-HMS-HMS4-XA-M40-M40B-T800HB)[36,39,55-57] ; pitch-based FT700 carbon fiber[50] ; PBZT(poly(p-phenylene benzobisthiazole)) fiber[58] ; pitch-based P75 carbon fiber[59]; Tyranno SiC fiber (UBE Ind., Jpn)[60] ; Nicalon SiC fiber[61-64].

**Fig. 6. (a)** Raman detection of carbon and SiC spectra as a function of the position along the fiber radius ($\lambda$=632.8nm). 100% corresponds to the maximum intensity detected for each phase ; **(b)** Bandwidths obtained after spectra fitting for "$sp^3$-carbon" peak ($\lambda$=632.8nm) and one of SiC TO modes ($\lambda$=514.5nm) ; **(c)** Young's modulus and "Berkovich's hardness" (after Mann et al[66]).
See in text what regions I to V correspond to.

**Table 1.** Properties of commercially available SiC fibers.

| Fiber grade | C/Si Atom | O (wt%) | Ret[♣] | <SiC>[ϖ] (nm) | SiC[<] (nm) | C (nm) | d (g.cm$^{-3}$) | E (GPa) | σ$_r$ (MPa) | ε$_r$ (%) |
|---|---|---|---|---|---|---|---|---|---|---|
| NLM[♦] | 1.22 [72] 1.34 [22] | 11.1 [72] | Ox | ~1 [72] 3 [18] 1.8 [79] 2 [80] 2.2 [32] | 1-4 [7] | ≤1[∅] [6] ~2 [79] | 2.56 [72] 2.55 [22] | 220 [72] 185 [3] 205 [22] 190 [33] | 2900 [22] 3000 [72] 2200±700 [3] 2970 [33] | 1.4 [22,72] 1.2±0.09 [3] |
| Hi | 1.39 [21] 1.41 [5] 1.55 [81] 1.35 [82] | 0.5 [21,33,72] | e- | 5 [5] 5.4 [32] 4.5 [21] | 2-15 [5,7] 5-20[$] [4] 3-5 [82] 1-30[&] [82] | 2-3[∂] [5] 2-5[⊗] [4] | 2.74 [4,33,72] 2.77 [5] 2.75 [21] | 270 [33,72] 300 [4] 261 [4] 263 [3] | 2800 [33,72] 2850 [4] 2710 [4] 2600±600 [3] | 1 [72] 0.95 [4] 0.9±0.03 [3] |
| Hi-S[32] | 1.05 | 0.2 | e- | 10.9 | / | / | 3.1 | 420 | ≥2500 | 0.6 |
| SA[*] | 1.08 [22] | ≈ 0.3 | Ox | 38 [22] 200 [21] | 100-300 [22] | # | 3.02 [21,22] | 420 [22] | 2800-3000 [22] | 0.7 [22] |
| Sylramic[33][♠] | ≈1 | 0.3 | / | / | 100-500 | / | 3.1 | 380 | 2950 | / |
| FT700 | / | / | / | / | / | / | 2.13 | 716 | 3600 | 0.5 |

[♣]Reticulation mode : Ox = thermal oxidation ; e- = electrons beam ; [ϖ]cited references give an average grain size (x-rays) ; [<]cited references give a size range based on electronic microscopy observations ; [♦]0.2 wt% H[72] ; [∅]2-3 layers of 10 "C$_6$-rings"[7] ; [$]mostly 5-7.5 nm ; [&]from Selected-Area Diffraction Pattern ; [∂]5-8 layers ; [⊗]5-10 layers ; [*]≤1wt% Al ; #At triple points[21] ; [♠]B = sintering aid ⇒ 3wt% TiB$_2$(0.5 µm)+1wt% B$_4$C(0.1 µm)

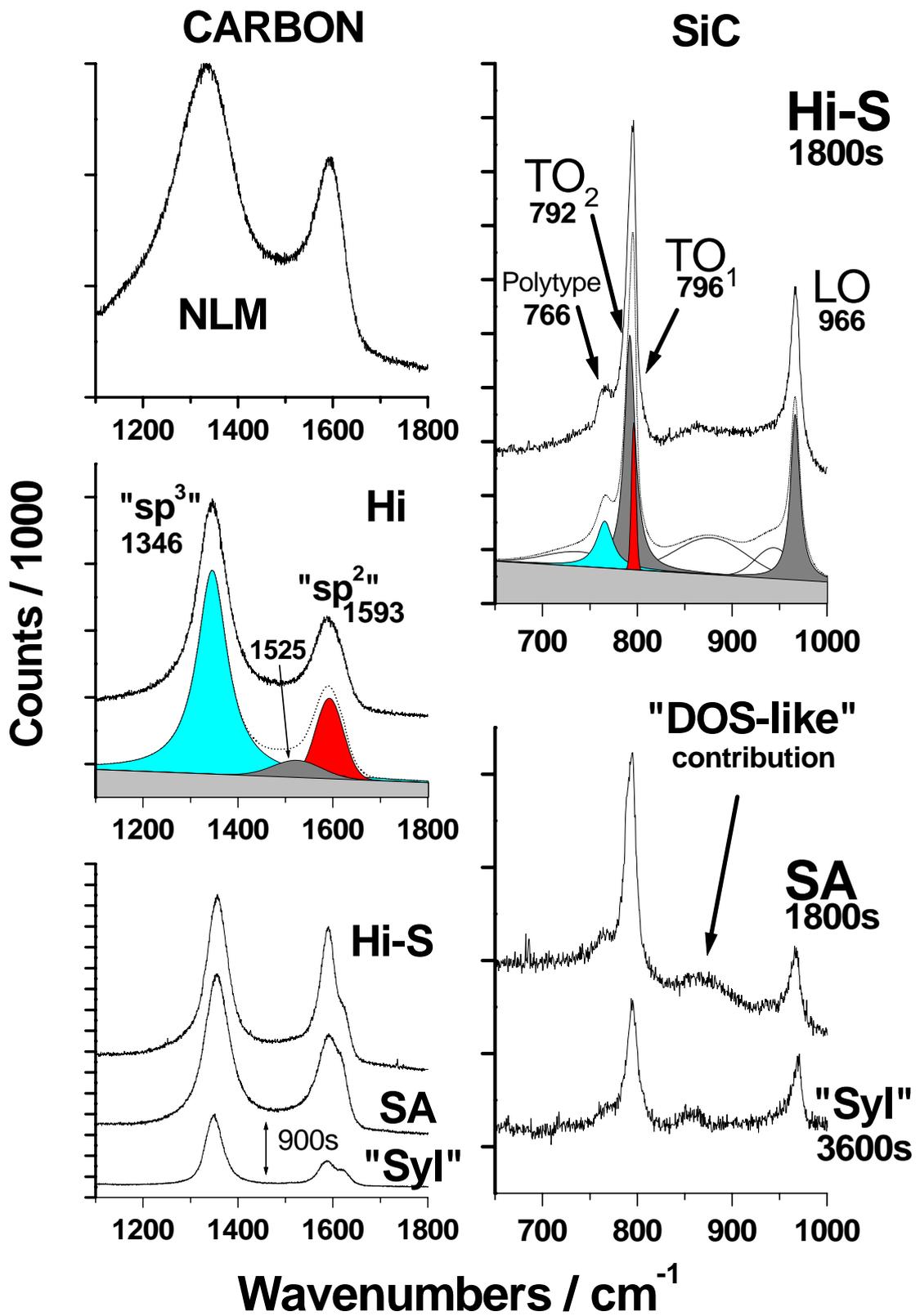

**Fig. 1.** Carbon (left) and SiC (right) spectra recorded on the successive generations of polymer-derived silicon carbide fibers ($\lambda$=514.5nm ; P=2mW ; t=90s, unless specified). One fitting is given in example in each case.

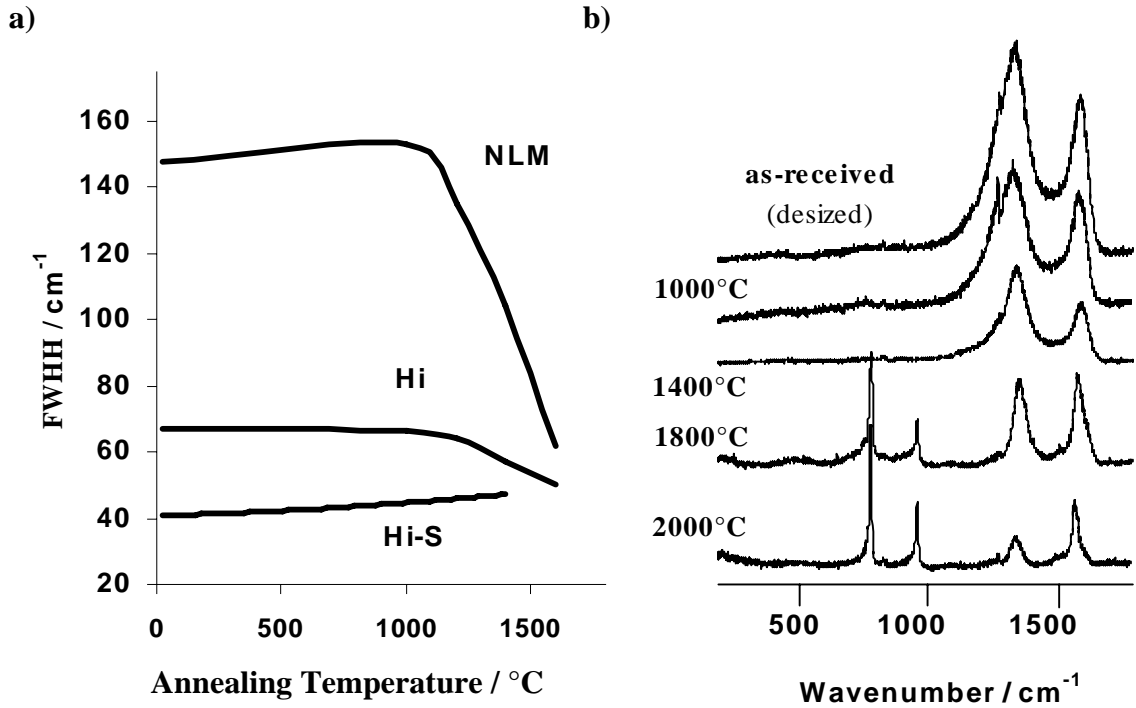

**Fig. 2.** **(a)** Full Width at Half Height (FWHH) of "sp$^3$" carbon mode as a function of thermal treatments performed in air for SiC fibers synthesized by oxygen (NLM) or electronic (Hi, Hi-S) curing of a polymeric precursor (polycarbosilane) ; **(b)** Changes of NLM core spectra on annealing in a reducing atmosphere.
Raman spectra in **(a)** and **(b)** were recorded with the 514.5nm line.

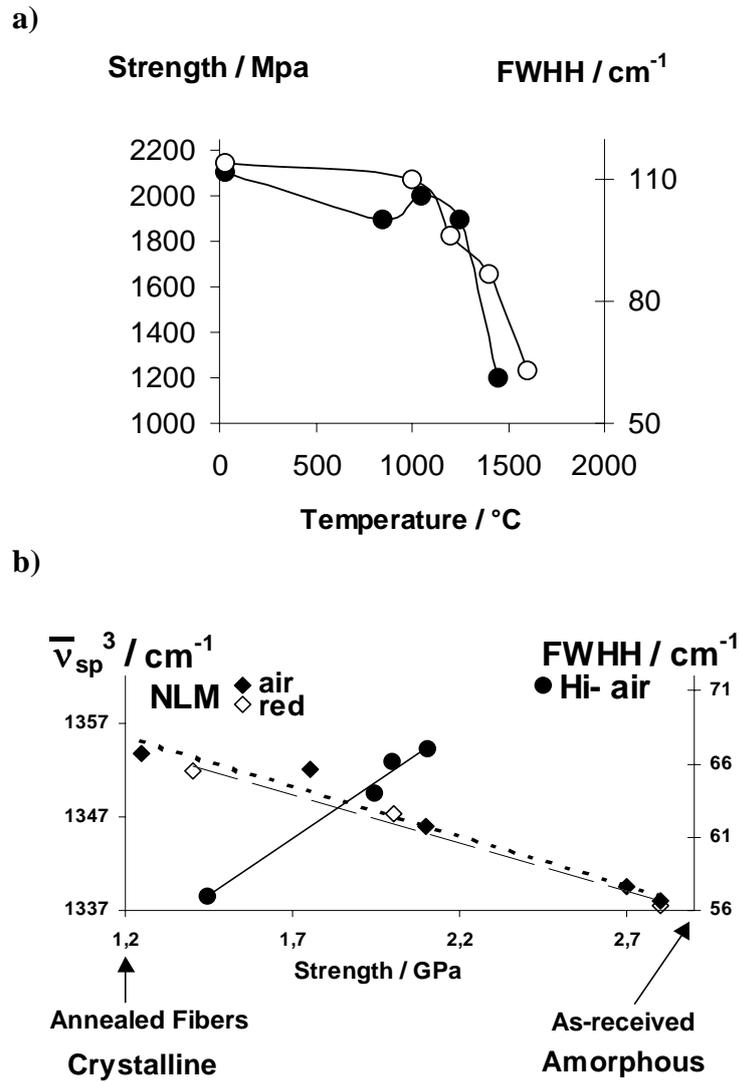

**Fig. 3.** (a) Comparison for different temperatures of Hi-Nicalon fibers tensile strength[21] (black circles) with the Full Width at Half Height (FWHH) of their "sp$^3$-carbon" stretching mode on surface recorded Raman spectra (white circles) ; (b) wavenumber and FWHH of "sp$^3$-carbon" mode plotted as a function of tensile strengths found in Berger et al[21] or Kumagawa et al[22] for heat treated fibers.
Raman spectra in (a) and (b) were recorded with the 514.5nm line.

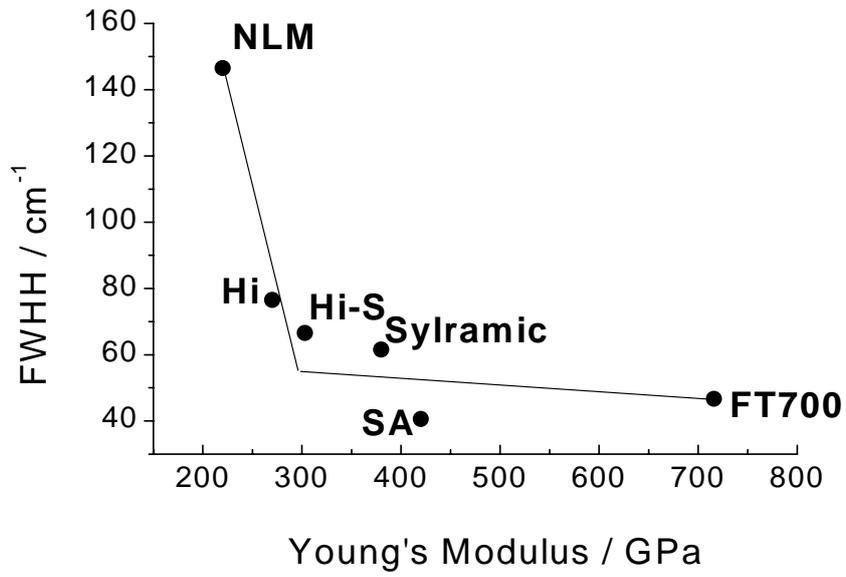

**Fig. 4.** Raman bandwidth ("sp$^3$" C-C mode) and Young's Modulus of as-received SiC fibers. FT700 carbon fiber (Textron, U.S.A.) is presented for comparison.

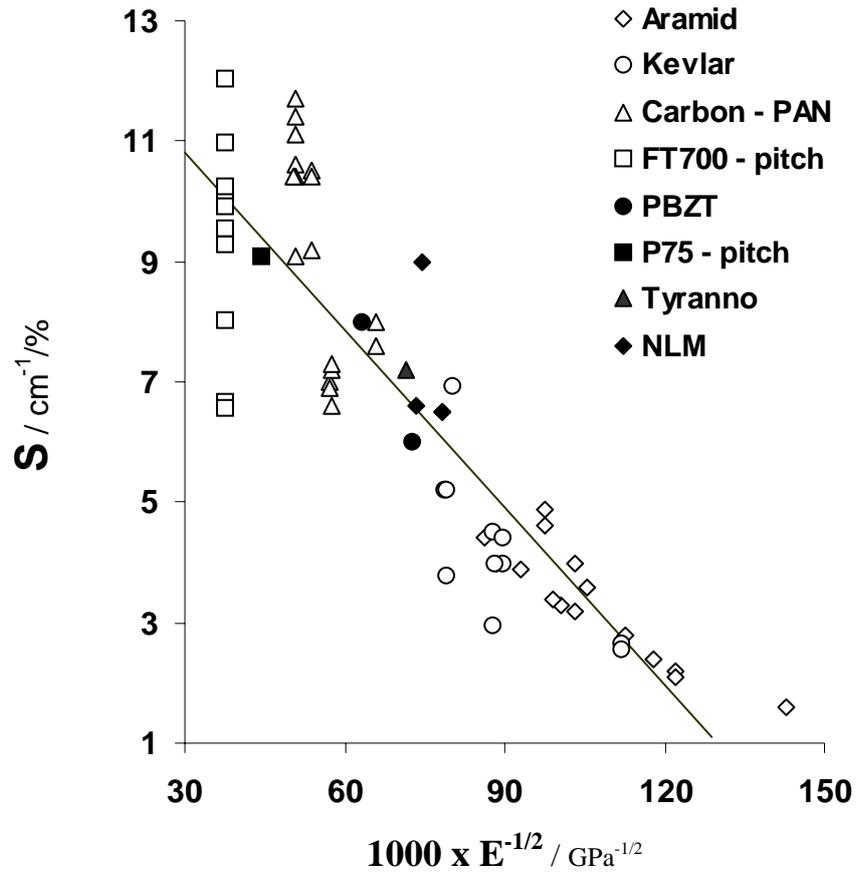

**Fig. 5.** Absolute value of the stress sensitivity coefficient (S) of carbon "sp$^2$" band plotted versus the inverse of Young's Modulus square root.

**References :** Kevlar™(poly(p-phenylene terephtalamide))[11,52-54] ; aramid fibers other than Kevlar™[11,40,51] ; PAN-based carbon fibers (IM43-HMS-HMS4-XA-M40-M40B-T800HB)[36,39,55-57] ; pitch-based FT700 carbon fiber[50] ; PBZT(poly(p-phenylene benzobisthiazole)) fiber[58] ; pitch-based P75 carbon fiber[59]; Tyranno SiC fiber (UBE Ind., Jpn)[60] ; Nicalon SiC fiber[61-64].

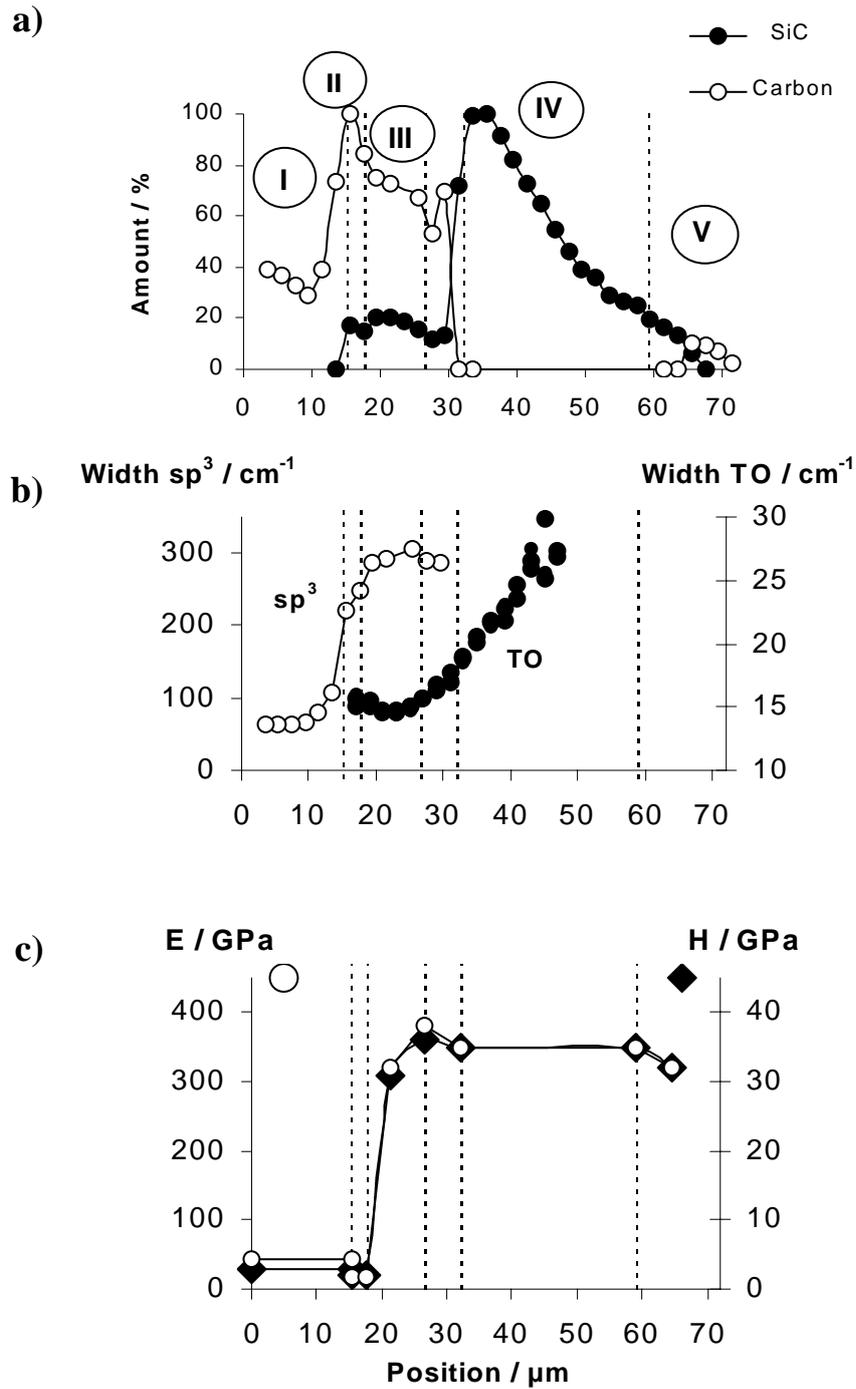

**Fig. 6. (a)** Raman detection of carbon and SiC spectra as a function of the position along the fiber radius ($\lambda$=632.8nm). 100% corresponds to the maximum intensity detected for each phase ; **(b)** Bandwidths obtained after spectra fitting for "sp$^3$-carbon" peak ($\lambda$=632.8nm) and one of SiC TO modes ($\lambda$=514.5nm) ; **(c)** Young's modulus and "Berkovich's hardness" (after Mann et al[66]).
See in text what regions I to V correspond to.

**References**

1. Chawla, K. K. *Composite Materials Science and Engineering, 2nd edition*, Springer-Verlag, New-York, 1998.
2. Karlin, S. & Colomban, Ph., Raman Study of the Chemical and Thermal Degradation of As-Received and Sol-Gel Embedded Nicalon and Hi-Nicalon SiC Fibres Used in Ceramic Matrix Composites. *J. Raman Spectrosc.,* 1997, **28**, 219-228.
3. Bunsell, A. R., Berger, M.-H. & Hochet, N., Structural and Mechanical Characterization of Some Alumina and SiC Based Fibres. *Ceramic Transactions,* 1995, **58**, 85-94.
4. Berger, M.-H., Hochet, N. & Bunsell, A. R., Microstructure and Thermo-Mechanical Stability of a Low-Oxygen Nicalon Fibre. *J. Microscopy,* 1995, **177**, 230-241.
5. Chollon, G., Pailler, R. & Naslain, R., Structure, Composition and Mechanical Behaviour at High Temperature of the Oxygen-Free Hi-Nicalon Fibre. *Ceramic Transactions,* 1995, **58**, 299-304.
6. Maniette, Y. & Oberlin, A., TEM Characterization of Some Crude or Air Heat-Treated SiC Nicalon Fibres. *J. Mater. Sci.,* 1989, **24**, 3361-3370.
7. Chollon, G., Pailler, R., Naslain, R. & Olry, P., Correlation Between Microstructure and Mechanical Behaviour at High Temperatures of a SiC Fibre with a Low Oxygen Content (Hi-Nicalon). *J. Mater. Sci.,* 1997, **32**, 1133-1147.
8. Lipowitz, J., Polymer-Derived Ceramic Fibers. *Ceram. Bull.,* 1991, **70**, 1888-1894.
9. Colomban, Ph., Gel Technology in Ceramics, Glass-Ceramics and Ceramic-Ceramic Composites. *Ceram. Inter.,* 1989, **15**, 23.
10. Colomban, Ph., Protonic Defects and Crystallization of Sol-Gel (Si,Ge) Mullites and Alumina. In *Ceramics Today-Tomorrow's Ceramics; Materials Science Monographs,* **66B**, Elsevier, Amsterdam, 1991, pp. 599-605.
11. Young, R. J., Lu, D., Day, R. J., Knoff, W. F. & Davis, H. A., Relationship between Structure and Mechanical Properties for Aramid Fibres. *J. Mater. Sci.,* 1992, **27**, 5431-5440.
12. Mouchon, E. & Colomban, Ph., Microwave Absorbent : Preparation, Mechanical Properties and R.F.-Microwave Conductivity of SiC (and/or Mullite) Fibre Reinforced Nasicon Matrix Composites. *J. Mater. Sci.,* 1996, **31**, 323-334.
13. Bansal, N. P., Effects of HF Treatments on Tensile Strength of Hi-Nicalon Fibres. *J. Mater. Sci.,* 1998, **33**, 4287-4295.
14. Carles, R., Mlayah, A., Amjoud, M. B., Reynes, A. & Morancho, R., Structural Characterization of Ge Microcrystals in $Ge_xC_{(1-x)}$ films. *Jap. J. Appl. Phys.,* 1992, **31** (pt 1), 3511-3514.
15. Gogotsi, Y. G., Kailer, A. & Nickel, K. G., Phase Transformations in Materials Studied by Micro-Raman Spectroscopy of Indentations. *Mater. Res. Innovat.,* 1997, **1**, 3-9.
16. Amer, M. S., Busbee, J., Leclair, S. R., Maguire, J. F., Johns, J. & Voevodin, A., Non Destructive, *In-situ* Measurements of Diamond-Like Carbon Film Hardness Using Raman and Rayleigh Scattering. *J. Raman. Spectrosc.,* 1999, **30**, 947-950.
17. Kailer, A., Nickel, K. G. & Gogotsi, Y. G., Raman Microspectroscopy of Nanocrystalline and Amorphous Phases in Hardness Indentations. *J. Raman Spectrosc.,* 1999, **30**, 939-946.
18. Sasaki, Y., Nishina, Y., Sato, M. & Okamura, K., Raman Study of SiC Fibres Made from PCS. *J. Mater. Sci.,* 1987, **22**, 443-448.

19. Gouadec, G., Karlin, S. & Colomban, Ph., Raman Extensometry Study of NLM202 and Hi-Nicalon SiC Fibres. *Composites,* 1998, **29B**, 251-261.
20. Gouadec, G., Colomban, Ph. & Bansal, N. P., Raman Study of Hi-Nicalon Fiber-Reinforced Celsian Composites, Part2 : Residual Stress in the Fibers. *J. Am. Ceram. Soc.,* accepted.
21. Berger, M.-H., Hochet, N. & Bunsell, A. R., Microstructure and High Temperature Mechanical Behavior of New Polymer Derived SiC Based Fibres. *Ceram. Eng. Sci. Proc.,* 1998, **19**, 39-46.
22. Kumagawa, K., Yamaoka, H., Shibuya, M. & Yamamura, T., Fabrication and Mechanical Properties of New Improved Si-M-C-(O) Tyranno Fiber. *Ceram. Eng. Sci. Proc.,* 1998, **19**, 65-72.
23. Gouadec, G., Karlin, S., Wu, J., Parlier, M. & Colomban, Ph., Physical Chemistry and Mechanical Imaging of Ceramic-Fibre-Reinforced Ceramic- or Metal-Matrix Composites. *Comp. Sci. Techn.,* 2000 (in press).
24. Gouadec, G. & Colomban, Ph., De l'Analyse Micro/Nano-Structurale et Micromécanique à l'Imagerie des Fibres de Renfort d'un Composite à Matrice Métallique (in french). *J. Phys. IV, Pr4,* 2000, **10**, 69-74.
25. Yoshikawa, M., Katagiri, G., Ishida, H., Ishitani, A. & Akamatsu, T., Resonant Raman Scattering of Diamondlike Amorphous Carbon Films. *Appl. Phys. Lett.,* 1988, **52**, 1639-1641.
26. Yoshikawa, M., Katagiri, G., Ishida, H., Ishitani, A. & Akamatsu, T., Raman Spectra of Diamond-Like Amorphous Carbon Films. *Solid State Comm.,* 1988, **66**, 1177-1180.
27. Wagner, J., Ramsteiner, M., Wild, C. & Koidl, P., Resonant Raman Scattering of Amorphous Carbon and Polycrystalline Diamond Films. *Phys. Rev. B,* 1989, **40**, 1817-1824.
28. Vidano, R. & Fischbach, D. B., New Lines in the Raman Spectra of Carbons and Graphite. *J. Am. Ceram. Soc.,* 1978, **61**, 13-17.
29. Marcus, B., Fayette, L., Mermoux, M., Abello, L. & Lucazeau, G., Analysis of the Structure of Multi-Component Carbon Films by Resonant Raman Scattering. *J. Appl. Phys.,* 1994, **76**, 3463-3470.
30. Dennison, J. R., Holtz, M. & Swain, G., Raman Spectroscopy of Carbon Materials. *Spectroscopy,* 1996, **11**, 38-46.
31. Gouadec, G., Colomban, Ph. & Bansal, N. P., Raman Study of Hi-Nicalon Fiber-Reinforced Celsian Composites, Part1 : Distribution and Nanostructure of Different Phases. *J. Am. Ceram. Soc.,* accepted.
32. Takeda, M., Sakamoto, J., Saeki, A., Imai, Y. & Ichikawa, H., High Performance Silicon Carbide Fibre Hi-Nicalon for CMCs. *Ceram. Eng. Sci. Proc.,* 1995, **16**, 37-44.
33. In *Product Information Form No 10-754-97*. Dow Corning Corp., Midland, Michigan, U.S.A., 1997.
34. Choyke, W. J. & Pensl, G., Physical Properties of SiC. *MRS Bull.,* 1997, **22**, 25-29.
35. Galiotis, C., Paipetis, A. & Marston, C., Unification of Fibre/Matrix Interfacial Measurements with Raman Microscopy. *J. Raman. Spectrosc.,* 1999, **30**, 899-912.
36. Paipetis, A. & Galiotis, C., A Study of the Stress-Transfer Characteristics in Model Composites as a Function of Material Processing, Fibre Sizing and Temperature of the Environment. *Comp. Sci. Tech.,* 1997, **57**, 827-838.
37. Young, R. J., Evaluation of Composite Interfaces Using Raman Spectroscopy. *Key Eng. Mat.,* 1996, **116-117**, 173-192.

38. Amer, M. S. & Schadler, L. S., The Effect of Interphase Toughness on Fibre/Fibre Interaction in Graphite/Epoxy Composites : An Experimental and Modelling Study. *J. Raman Spectrosc.,* 1999, **30**, 919-928.
39. Lévêque, D. & Auvray, M. H., Study of Carbon-Fibre Strain in Model Composites by Raman Spectroscopy. *Comp. Sci. Tech.,* 1996, **56**, 749-754.
40. Bennett, J. A. & Young, R. J., Micromechanical Aspects of Fibre/Crack Interactions in an Aramid/Epoxy Composite. *Comp. Sci. Tech.,* 1997, **57**, 945-956.
41. Arahori, T., Iwamoto, N. & Umesaki, N., Influence of $Y_2O_3$ Addition on the Transformation Behavior of $ZrO_2$ in $Al_2O_3$-$ZrO_2$ Composites. *J. Ceram. Soc. Jpn. Inter. Ed.,* 1987, **95**, 898-904.
42. Alzyab, B., Perry, C. H. & Ingel, R. P., High-Pressure Phase Transitions in Zirconia and Yttria-Doped Zirconia. *J. Am. Ceram. Soc.,* 1987, **70**, 760-765.
43. Provoost, R., Rosseel, K., Moshchalkov, V. V., Silverans, R. E., Bruynseraede, Y., Dierickx, D. & Van-der-Biest, O., Stress Release at $Y_2BaCuO_5$ Inclusions in Fast Melt Processed $YBa_2Cu_3O_{7-x}$ Observed by Micro-Raman Spectroscopy. *Appl. Phys. Lett.,* 1997, **70**, 2897-2899.
44. Wu, J. & Colomban, Ph., Raman Spectroscopy Study on the Stress Distribution in the Continuous Fibre-Reinforced CMC. *J. Raman Spectrosc.,* 1997, **28**, 523-529.
45. Karlin, S. & Colomban, Ph., Micro Raman Study of SiC-Oxide Matrix Reaction. *Composites,* 1998, **29B**, 41-50.
46. Yang, X. & Young, R. J., The Microstructure of a Nicalon/SiC Composite and Fibre Deformation in the Composite. *J. Mater. Sci.,* 1993, **28**, 2536-2544.
47. Chollon, G. & Takahashi, J., La Microscopie Raman Appliquée aux Composites Carbone/Carbone (in french). In *Proceedings of JNC11 (Journées Nationales sur les Composites)*, eds J. Lamon & D. Baptiste. AMAC, Paris, 1998, vol.**2**, pp. 777-785.
48. White, M. A., Thermal Properties of Solids : Etude in Three-part Anharmony. *Can. J. of Chem.,* 1996, **74**, 1916-1921.
49. El-Mallawany, R. & Abd-El-Moneim, A., Comparison between the Elastic Moduli of Tellurite and Phosphate Glasses. *Phys. State Sol. (a),* 1998, **166**, 829-834.
50. Gouadec, G. & Colomban, Ph., Raman Extensometry : Anharmonicity and Stress (in french). In *Proceedings of JNC11 (Journées Nationales sur les Composites)*, eds J. Lamon & D. Baptiste. AMAC, Paris, 1998, vol.**2**, pp. 759-766.
51. Van der Zwaag, S., Northolt, M. G., Young, R. J., Robinson, I. M., Galiotis, C. & Batchelder, D. N., Chain Stretching in Aramid Fibres. *Polym. Comm.,* 1987, **28**, 276-277.
52. Vlattas, C. & Galiotis, C., Monitoring the Behavior of Polymer Fibres Under Axial Compression. *Polymer,* 1991, **32**, 1788-1793.
53. Galiotis, C., Laser Raman Spectroscopy, a New Stress/Strain Measurement Technique for the Remote and On-line Nondestructive Inspection of Fiber Reinforced Polymer Composites. *Mater. Techn.,* 1993, **8**, 203-209.
54. Arjyal, B. & Galiotis, C., Application of a Laser Raman Sensor for Stress Monitoring in Composites. In *Proceedings of 3rd Inter. Conf. on Intelligent Materials (ICIM),* eds P.F. Gobin & J. Tatibouët. SPIE, Bellingham, Wash., U.S.A., 1996, vol. **2779**, pp. 142-145.
55. Melanitis, N. & Galiotis, C., Compressional Behavior of Carbon Fibres. Part 1 : a Raman Spectroscopy Study. *J. Mater. Sci.,* 1990, **25**, 5081-5090.

56. Beyerlein, I. J., Amer, M. S., Schadler, L. S. & Phoenix, S. L., New Methodology for Determining *in-situ* Fibre, Matrix and Interfaces Stresses in Damaged Multifiber Composites. *Sci. Eng. Comp. Mater.,* 1998, **151**, 204.
57. Young, R. J. & Day, R. J., Application of Raman Microscopy to the Analysis of High Modulus Polymer Fibres and Composites. *Brit. Polymer J.,* 1989, **21**, 17-21.
58. Day, R. J., Robinson, I. M., Zakikhani, M. & Young, R. J., Raman Spectroscopy of Stressed High Modulus Poly(p-phenylene benzobisthiazole) Fibres. *Polymer,* 1987, **28**, 1833-1840.
59. Filiou, C., Galiotis, C. & Batchelder, D. N., Residual Stress Distribution in Carbon Fibre/Thermoplastic Matrix Pre-Impregnated Composite Tapes. *Composites,* 1992, **23**, 28-38.
60. Bollet, F., Galiotis, C. & Reece, M. J., Characterization of Strain Distribution in Fibres Bridging Ceramic Matrix Cracks by LRS. In *Proceedings of $7^{th}$ Eur. Conf. on Comp. Mater (ECCM-7)*. Woodhead Publishing Ltd, Cambridge, England, 1996, vol. **1**, pp. 505-510.
61. Yang, X. & Young, R. J., Model Ceramic Fibre-Reinforced Glass Composites : Residual Thermal Stresses. *Composites,* 1994, **25**, 488-493.
62. Yang, X., Bannister, D. J. & Young, R. J., Analysis of the Single-Fiber Pullout Test Using Raman Spectroscopy : Part III, Pullout of Nicalon Fibers from a Pyrex Matrix. *J. Am. Ceram. Soc.,* 1996, **79**, 1868-1874.
63. Day, R. J., Piddock, V., Taylor, R., Young, R. J. & Zakikhani, M., The Distribution of Graphitic Microcrystals and the Sensitivity of their Raman Bands to Strain in SiC Fibres. *J. Mater. Sci.,* 1989, **24**, 2898-2902.
64. Yang, X. & Young, R. J., Fibre Deformation and Residual Strain in Silicon Carbide Fibre Reinforced Glass Composites. *Brit. Ceram. Trans.,* 1994, **93**, 1-10.
65. Ning, X. J. & Pirouz, P., The Microstructure of SCS-6 SiC Fibre. *J. Mater. Res.,* 1991, **6**, 2234-2248.
66. Mann, A. B., Balooch, M., Kinney, J. H. & Weihs, T. P., Radial Variations in Modulus and Hardness in SCS-6 Silicon Carbide Fibers. *J. Am. Ceram. Soc.,* 1999, **82**, 111-116.
67. Gilkes, K. W. R., Sands, H. S., Batchelder, D. N., Robertson, J. & Milne, W. I., Direct Observation of $sp^3$ Bonding in Tetrahedral Amorphous Carbon using UV Raman Spectroscopy. *Appl. Phys. Lett.,* 1997, **70**, 1980-1982.
68. Wan, J. Z., Pollak, F. H. & Dorfman, B. E., Micro-Raman Study of Diamondlike Atomic-Scale Composite Films Modified by Continuous Wave Laser Annealing. *J. Appl. Phys.,* 1997, **81**, 6407-6414.
69. Colomban, Ph., Raman Microspectrometry and Imaging of Ceramic Fibers in CMCs and MMCs. In *Advances in Ceramic Matrix Composites V*, American Ceramic Society, Westerville, OH, U.S.A., 2000, pp. 517-540.
70. Laffon, C., Flank, A. M., Lagarde, P., Laridjani, M., Hagege, R., Olry, P., Cotteret, J., Dixmier, J., Miquel, J. L., Hommel, H. & Legrand, A. P., Study of Nicalon-Based Ceramic Fibres and Powders by EXAFS Spectrometry, X-ray Diffractometry and some Additional Methods. *J. Mater. Sci.,* 1989, **24**, 1503-1512.
71. Nakashima, S. & Harima, H., Raman Investigation of SiC Polytypes. *Phys. Stat. Sol. (a),* 1997, **162**, 39-64.
72. Iskikawa, T., Recent Developments of the SiC Fibre Nicalon and its Composites, Including Properties of the SiC fibre Hi-Nicalon for Ultra-High Temperature. *Comp. Sci. Techn.,* 1994, **51**, 135-144.

73. Schreck, P., Vix-Guterl, C., Ehrburger, P. & Lahaye, J., Reactivity and Molecular Structure of Silicon Carbide Fibres Derived from Polycarbosilanes. Part1 Thermal Behavior and Reactivity. *J. Mater. Sci.,* 1992, **27**, 4237-4242.
74. Sacks, M. D., Effect of Composition and Heat Treatment Conditions on the Tensile Strength and Creep Resistance of SiC-based Fibers. *J. Europ. Ceram. Soc.,* 1999, **19**, 2305-2315.
75. Shimoo, T., Toyoda, F. & Okamura, K., Thermal Stabilty and Low-Oxygen Silicon Carbide Fiber (Hi-Nicalon) Subjected to Selected Oxidation Treatment. *J. Am. Ceram. Soc.,* 2000, **83**, 1450-1456.
76. Ishikawa, T., Kohtoku, Y., Kumagawa, K., Yamamura, T. & Nagasawa, T., High-Strength Alkali-Resistant Sintered SiC Fibre Stable to 2,200°C. *Nature,* 1998, **391**, 773-775.
77. Tarantili, P. A., Andreopoulos, A. G. & Galiotis, C., Real-Time Micro-Raman Measurements on Stressed Polyethylene Fibers. 1. Strain Rate Effects and Molecular Stress Redistribution. *Macromolecules,* 1998, **31**, 6964-6976.
78. Melanitis, N., Tetlow, P. L., Galiotis, C. & Smith, S. B., Compressional Behaviour of Carbon Fibres, Part II : Modulus Softening. *J. Mater. Sci.,* 1994, **29**, 786-799.
79. Chauvet, O., Stoto, T. & Zuppiroli, L., Hopping Conduction in a Nanometer-Size Crystalline System : a SiC Fiber. *Phys. Rev. B,* 1992, **46**, 8139-8146.
80. Vahlas, C., Rocabois, P. & Bernard, C., Thermal Degradation Mechanisms of Nicalon Fibre : a Thermodynamic Simulation. *J. Mater. Sci,* 1994, **29**, 5839-5846.
81. Bodet, R. & Lamon, J., Comportement en Fluage de Fibres Céramiques à Base SiC. *Silicates Industriels,* 1996, **1-2**, 23-31.
82. Bansal, N. P. & Chen, Y. L., Chemical, Mechanical and Microstructural Characterization of Low-Oxygen Containing Silicon Carbide Fibers with Ceramic Coatings. *J. Mater. Sci.,* 1998, **33**, 5277-5289.